\documentclass[usenatbib]{mn2e}
\usepackage{times}
\usepackage{epsfig}
\usepackage{amsmath}
\title[The hierarchical buildup of stars]{Testing Cold Dark Matter with
  the hierarchical buildup of stellar light}
\author[Balogh et al.]{Michael L. Balogh$^{1}$, Ian G. McCarthy$^{2}$, Richard
  G. Bower$^{2}$, Vincent R. Eke$^{2}$\\
$^{1}$Department of Physics and Astronomy, University of Waterloo, Waterloo, Ontario, N2L 3G1, Canada\\
$^{2}$Department of Physics, University of Durham, Durham, UK, DH1 3LE\\
}

\date{\today}

\def\gtrsim{\mathrel{\raise0.35ex\hbox{$\scriptstyle >$}\kern-0.6em
\lower0.40ex\hbox{{$\scriptstyle \sim$}}}}
\def\lesssim{\mathrel{\raise0.35ex\hbox{$\scriptstyle <$}\kern-0.6em
\lower0.40ex\hbox{{$\scriptstyle \sim$}}}}

\def\kms{km~s$^{-1}$}
\def\r500{$R_{500}$}
\def\m500{$M_{500}$}
\def\mstar{$M_{\ast,500}$}
\def\fstar{$f_{\ast}$}
\def\b{d\log{f_\ast}/d\log{M_{500}}}
\begin{document} 
\maketitle
\begin{abstract}
The hierarchical growth of mass in the Universe is a pillar of all cold
dark matter (CDM) models.   In this paper we demonstrate that this principle
leads to a robust, falsifiable prediction of the stellar content of
groups and clusters, that is testable with current observations and
is relatively insensitive to the details of baryonic physics or cosmological parameters.  Since it is difficult to
preferentially remove stars from dark-matter dominated systems, when
these systems merge the fraction of total mass in stars can only increase
(via star formation) or remain constant, relative to the fraction
in the combined systems prior to the merger.  Therefore, hierarchical
models can put strong constraints on the observed correlation between
stellar fraction, \fstar, and total system mass, $M_{500}$.  In particular,
if this relation is fixed and does not evolve with redshift, CDM models
predict $b=\b\gtrsim-0.3$.  This constraint can be weakened if the
\fstar--\m500\ relation evolves strongly, but this
implies more stars must be formed in situ in groups at low redshift.
Conservatively requiring that at least half the stars in groups were
formed by $z=1$, the constraint from evolution models is $b\gtrsim-0.35$.  Since the most massive clusters
($M\sim10^{15}M_\odot$) are observed to have \fstar$\sim 0.01$, this means that groups
with $M=5\times10^{13}M_\odot$ must have \fstar$\leq0.03$.  Recent
observations by \citet{GZZ} indicate a much steeper relation, with
\fstar$>0.04$ in groups leading to $b\approx -0.64$.  If
confirmed, this would rule out hierarchical structure formation
models: today's clusters could not have been built from today's groups,
or even from the higher-redshift progenitors of those groups.  We
perform a careful analysis of these and other data to identify the
most important systematic uncertainties in their measurements.
Although correlated uncertainties on stellar and total masses might
explain the steep observed relation, the data are only consistent with
theory if the observed group masses are systematically underestimated.
\end{abstract}
\section{Introduction}\label{sec-intro}
An inescapable prediction of all cold dark matter (CDM) models is that mass
in the Universe, 
in the form of CDM dominated ``haloes'', builds up {\it hierarchically}, with low-mass systems
merging to form progressively more massive galaxies and clusters of
galaxies \citep[e.g.][]{BBKS,DEFW}.  Moreover, the rate of this mass growth is precisely
determined for a given set of cosmological parameters \citep[e.g.][]{GW07}.  In practice, the complex and non-linear nature of baryonic
physics \citep[particularly cooling and heating processes, ][and many others]{WR78,WF91,Cole2000,bower06,Croton05} means that this is
a difficult prediction to test through observations of galaxies alone.
In fact, observations show that galaxies form in the
opposite way, with the most massive systems having their stars in place
first \citep[e.g.][]{Cowie+96,Juneau+04,Pozzetti+07}.  This is not
considered a falsification of the cold dark matter model, because the
effect can be qualitatively
explained by improving the physical description of baryonic processes
in the models \citep[e.g.][]{Croton05,bower06}, in a way that leaves
untouched the
prediction of hierarchical growth in the dark matter component.  

However, an interesting and robust test of the theory can be obtained from
observations of the stellar fraction of dark-matter dominated
structures.  Unlike gas, it is very difficult to separate stars from dark matter, since
they are both collisionless forms of matter that interact only via
gravity.  And while new stars can be formed from gas (a process which
is very poorly understood) they can only be destroyed through normal
stellar evolution processes (which are quite well understood).  The
latter effect only removes 10-30 per cent of the total stellar mass, with this range
reflecting a weak dependence on star formation history and initial mass
function \citep[e.g.][]{JCP,BC03}.  Therefore,  when two similar systems merge, the mass fraction in visible
stars must be at least as
large as the fraction in the combined system prior to the merger, and
the simplest expectation is that \fstar\ will either be constant or
increase with total system mass.  

Stellar fractions are most reliably measured for galaxy clusters,
where the total mass in dark matter can be determined in various,
independent ways (e.g. gravitational lensing, X-ray gas, or galaxy
dynamics).  Interestingly, numerous studies have consistently shown
that \fstar\ of clusters and groups {\it decreases} with
increasing mass
\citep[e.g.][]{H+00,MH02,Eke-groups3,G+02,RBGMR,LMS,cnoc2_ir}.  The
usual explanation is that clusters are built not only from groups, but
also from the accretion of low mass galaxies, where 
\fstar\ must be very low to explain the
shallow faint-end slope of the luminosity function \citep[e.g.][]{WF91,MH02}.  Another
  possibility is that low-mass groups form a significant number of
  stars, but only {\it after} most clusters have been assembled.   
Either possibility allows theory to accommodate a
mildly decreasing \fstar\ on cluster scales; it is our goal in this
paper to use conservative constraints on these effects to put a robust limit on just how steep this decrease can be.

An important omission in many of the observational studies above, however,  has been the contribution
from intracluster light (ICL), a low-surface brightness distribution of
stars in groups and clusters that is very difficult to measure.
Most studies of rich clusters find that the ICL contribution is
relatively small, contributing less than 30 per cent to the total
stellar light \citep[e.g.][]{Durrell+02,Feld+04,Covone+06,KB07}, with
at most a weak dependence on system mass \citep{Zibetti}.
Recently, using deep $I-$band observations of 23 nearby systems,
\citet{GZZ1} have made careful measurements of the ICL
component, and come to the surprising conclusion that both \fstar\ and
the relative ICL contribution depends
much more strongly on mass than has been found previously, with the ICL
actually
dominating the total stellar mass in groups
\citep[][hereafter GZZ]{GZZ}.
This has motivated us to consider whether or not these
observations are able to falsify the hierarchical structure growth model.

We will begin by reanalyzing the observational data of
GZZ, 
and
complementary data from
\citet{LM04} in \S~\ref{sec-data}.  
In \S~\ref{sec-models} we use theoretical predictions for the growth of
dark matter structure to put robust, falsifiable limits on the
mass-dependence of \fstar.   This prediction is
then directly confronted
with the observational data in
\S~\ref{sec-test}, where we also discuss the implications of our findings, and the effect of
possible biases and uncertainties in the measurements.
Throughout this paper we generally assume a
cosmology with $\Omega_m=0.3$, $\Omega_\Lambda=0.7$, and
$H_\circ=70$\kms.  However we explicitly consider how our results
depend on cosmological parameters, in \S~\ref{sec-cosmo}.

\section{The stellar fraction in local clusters and groups}\label{sec-data}
\subsection{Description of the data}
We require an accurate account of the relative stellar
content for a fair sample of galaxy clusters and groups. 
We will take most of our data from two of the best recent surveys, GZZ
and \citet{LM04},
which are generally complementary in their sources of systematic uncertainty. 

GZZ 
have measured the total light in galaxies and
intracluster light, for 23 nearby clusters and groups.  They use
drift-scan observations with careful attention to flat-fielding, and
fit a two-component \citet{dV61} profile to the brightest cluster
galaxy (BCG).  The outer
component, with scale lengths of typically a few hundred kpc, is
interpreted as the ICL.  Stellar masses are obtained from the
integrated $I-$band luminosity, assuming a mass-to-light ratio
of $M/L_I=3.6$, based on
dynamically--determined masses for elliptical galaxies
\citep{Capp+06}.   These masses include a small (about 30\%)
contribution from the dark matter component; thus we adopt a stellar
$M/L=2.8$ for our analysis, in good agreement with stellar population
models assuming a \citet{Kroupa} initial mass function, as described in
\citet{Capp+06}.  The exact value used does not impact our
conclusions, which are derived from the trend of the stellar
fraction with system mass, rather than the normalization.

For these 23 clusters, the total $I-$ band light from the BCG and ICL are well
characterized, though the relative contribution of each cannot be so
robustly determined.  The measurement of the total galaxy light
is made with a statistical subtraction of the foreground and background population.
The main statistical uncertainty in these data arises from the total cluster masses, which are estimated from the line-of-sight velocity
dispersion of the galaxies, $\sigma$.
The relationship between velocity dispersion and dynamical mass is sensitive
to the total potential shape, and velocity anisotropy.  GZZ partially 
alleviate this uncertainty by employing a calibration between $\sigma$ and
X-ray derived masses, from an independent cluster sample
\citep{Vik+06}.  However, this
calibration sample is very small (13 clusters), so it is not possible
to determine the scatter about the mean relationship.  Moreover,
the four lowest mass systems in the GZZ sample require extrapolations of
this calibrating relationship, and may therefore be the most unreliable.

The second sample we consider is that of Lin, Mohr \& Stanford (2004)\nocite{LMS} and
\citet[][hereafter LM]{LM04},
who analyze $K-$band observations of 93 X-ray selected clusters, using
the 2 Micron All Sky Survey \citep[2MASS,][]{2MASS}.
This has the
advantages that the stellar mass-to-light ratio ($M/L$) is only weakly
dependent on star formation history, and
that the total system masses can be determined from the X-ray
temperatures.  To ensure the stellar masses can be fairly compared with
those of GZZ, we choose an average $M/L_K=0.9$; this is consistent
with the value of $M/L_I=2.8$ adopted for the GZZ data if $I-K=2.0$, a
reasonable number for the early-type galaxies expected to dominate
these clusters \citep[e.g.][]{P97,Smail01,KL05,Coma_phot}.  The $M/L_K$ we adopt
is somewhat higher than the average value used by \citet{LMS} 
but, again, the absolute value is of little consequence for our analysis.
We update the total mass estimates in LM using accurate {\it ASCA}
temperatures available for 63 of their clusters
 \citep{Hornerthesis,HMS}.  These temperatures are converted  into
 $M_{500}$\footnote{\m500\ is defined as the total mass within a radius
  \r500, such that the average density within \r500\ is 500 times the
  critical density at the cluster redshift.} using the relation found by \citet{Vik+06}, where $M_{500}$
 was determined from {\it Chandra} resolved surface brightness and temperature
 profiles.   This typically results in a $\sim 20$ per cent change to
 the mass, usually in the sense that our new masses are larger.  This
 corresponds to a $\sim 6$ per cent change to \r500; therefore we must also
 correct the measurement of total stellar light within this radius.
 We simply assume the galaxies follow a \citet{NFW} profile,
 with a scale radius $r_s=R_{500}/3$, and adjust the total stellar
 light measured by LM accordingly.  All of these corrections are small,
 and do not influence our conclusions at all; however, the more precise
 temperature measurements will result in a more precise mass estimate
 which, as we discuss in \S~\ref{sec-correrr}, is relevant to our
 interpretation of the data.

\begin{figure}
\leavevmode \epsfysize=8cm \epsfbox{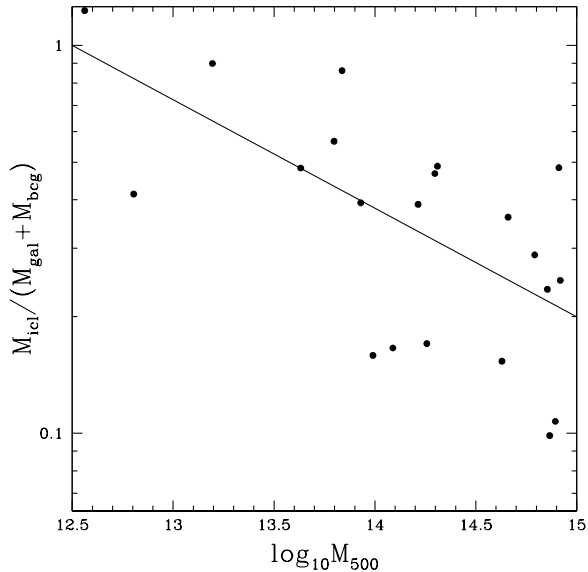}
\caption{For the cluster sample of \citet{GZZ1}, we show the amount of
  stellar mass in the intracluster light, relative to that in the
  galaxies (including the BCG), as a function of total mass \m500.  The
  solid line shows the relation we adopt to correct the data of
  LM for the intracluster light component.
\label{fig-iclfrac}}
\end{figure}

Unfortunately, the infrared data of LM are not deep enough to measure the
ICL contribution directly.  
To make use of these data we
need to make an approximate correction for this missing light, and we will use the
data of \citet{GZZ1} for this purpose.  In Figure~\ref{fig-iclfrac} we
show the mass of the ICL, $M_{\rm icl}$, relative to the
mass in galaxies (including the BCG), as a function of
\m500.  There is
a correlation (although largely driven by the four poorly calibrated, lowest-mass
systems) and we find approximately 
\begin{equation}
\log{\left[\frac{M_{\rm icl}}{M_{\rm gal}+M_{\rm
      bcg}}\right]}=-0.28\left(\log{M_{500}}-12.5\right),
\end{equation}
shown as the solid line.  That is, for the most massive clusters the
ICL contributes another 20\% to the mass observed in galaxies, while for
the lowest mass systems the total stellar mass is approximately doubled by
including the ICL.  We will apply this correction to the data of
LM.   It should be kept in mind that the ICL and BCG light
are not robustly separated by \citet{GZZ1}, and the BCG light in
particular is not measured the same way in both studies.  Nonetheless, the
correction is not likely to be grossly incorrect, and small differences
will not change our main conclusions (see further discussion in
\S~\ref{sec-obserrs}).  

\subsection{Correlated Uncertainties}\label{sec-correrr}
An important consideration in this analysis is proper accounting for
uncertainties in the measured quantities.  In both GZZ and LM, the
statistical uncertainty on \mstar\ is small, approximately $10$ per
cent.  The dominant statistical uncertainty is in \m500,
particularly for the GZZ data, where \m500\ is derived from velocity
dispersions.   While the
number of redshifts per cluster in the GZZ sample is generally more than 20, the
uncertainties on $\sigma$ are still typically $\sim 10$\%, which
translates to a $\sim 30$\% uncertainty on mass, since \m500$\propto\sigma^3$.
The LM data generally have smaller
statistical uncertainties on the masses, partly due to the fact that
the mass dependence on temperature (\m500$\propto T^{1.5}$) is weaker than
on $\sigma$.  

However, the error analysis is more complex than this because \mstar\ is the total stellar mass measured within a
radius \r500, which depends on \m500.  Therefore, a statistical
overestimate of \m500\ will also result in an overestimate of \mstar,
by an amount that depends on the radial profile of the stellar mass distribution.
For the GZZ sample, we use the published radial profile of the BCG and ICL
component from \citet{GZZ1}.  For the galaxy component, we only know
the total mass within \r500.  We will therefore assume the galaxy mass
follows a \citet{NFW} profile, with a scale radius $r_s=R_{500}/3$.
For each cluster we can then directly calculate
$dM_{\ast,500}/dM_{500}$ and therefore the correlated uncertainty on
\mstar, $\Delta M_{\ast,500}=\left(dM_{\ast,500}/dM_{500}\right)\Delta{M_{500}}$.
For the LM data, we do not know the shape of the stellar mass profiles,
so we will simply adopt the average value of $dM_{\ast,500}/dM_{500}$
from the GZZ data; in any case the error bars for these data are
generally smaller than the data points in our figures, so this is of no
consequence.

\subsection{The stellar fraction in the most massive clusters}\label{sec-clufstar}
The most massive clusters are the systems for which all matter is
most reliably accounted for observationally.  The hot gas is visible as
X-ray emission, with a temperature closely related to the gravitational
potential, and the total system mass can be directly measured via
weak-lensing; both of these methods yield masses in good agreement with those
estimated from galaxy velocity dispersions \citep[e.g.][]{HEHY}.  Moreover, the stellar component is
dominated by old, passive galaxies at the present day, so k-corrections
and stellar $M/L$ ratios are relatively well determined. 

Considering just the most massive systems in 
GZZ, 
those with
$M_{500}>3\times10^{14}M_\odot$, there are eight clusters with stellar
fractions ranging from \fstar$=0.006$ to 0.02, and a median of 0.011.
In this same mass range, LM find that
the fraction of mass in {\it galaxies} spans a range of 0.005 to 0.024,
with a median of about 0.0095. Including a 20\% correction for
intracluster light, appropriate for these systems (see
\S~\ref{sec-data}), brings the median stellar fraction to 0.011, in excellent
agreement with the result of GZZ.

Although these two studies probably provide the most robust
measurement of the stellar mass fraction in clusters, the results are
consistent with those of many other studies.  For example,
\citet{Eke-groups2} find a $B-$band mass to light ratio of $\sim 350$
for the most massive systems in the 2dFGRS.  Assuming a typical (but
very model-dependent) stellar $M/L_B=4.5$ \citep{FHP}, this corresponds
to a stellar fraction in clusters of 0.013.  \citet{G+02} find a
higher
value of 0.026, assuming the same stellar mass-to-light ratio.  Most studies
tend to find values between these two; see references within
\citet{Eke-groups2} and \citet{G+02} for a good compilation.

The evidence is therefore good that the stellar fraction in clusters is
about 1\% on average and certainly $<3$\%.  
We note that this is very similar to the global stellar fraction
measured from large redshift surveys. \citet{Eke-groups3} combined the
2dFGRS \citep{2dF_colless} and 2MASS \citep{2MASS} surveys
to find an overall stellar fraction of 0.016 \cite[assuming a][initial mass function]{K83}, and this is consistent with
recent results from many other studies \citep[e.g.][and references within]{GBCW}.
This small number is
already known to put strong
constraints on the efficiency of galaxy formation when combined with
measurements of the hot gas mass \citep[e.g.][]{baryons}.  In
\S~\ref{sec-models} we will
show that the stellar fraction alone, independent of how much hot gas
may be present, can be used to test models of hierarchical
structure formation.

\subsection{Observed Mass-Dependence of \fstar}\label{sec-fstar}
\begin{figure}
\leavevmode \epsfysize=8cm \epsfbox{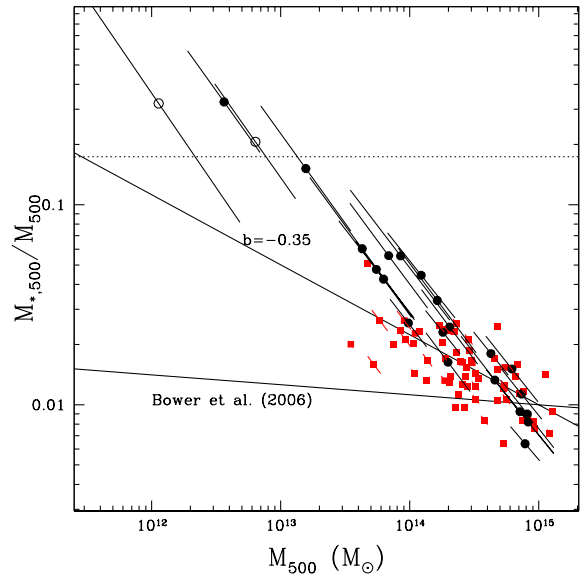}
\caption{The stellar mass fraction \fstar$=M_{\ast,500}/M_{500}$,
  including intracluster light,  is shown as a function of total mass
  \m500.  The LM data (red squares) include a correction for ICL
  estimated from the GZZ data (circles).  The two open circles
  represent clusters  A2405 and APMC020, which are systems strongly
  affected by line-of-sight structure.  The $1\sigma$ error
  bars are derived from the published uncertainties on \m500, and the
  tilt reflects the correlated uncertainty in \m500\ and \mstar$/$\m500
  as described in \S~\ref{sec-correrr}.  
  The horizontal, dotted line shows the global
  baryon fraction measured by WMAP3 \citep{WMAP3}.  The two solid lines
  show constant slopes of $-0.35$ and $-0.05$, for comparison with our most
  conservative theoretical lower limit, and the \citet{bower06} model prediction, respectively.
\label{fig-fstar}}
\end{figure}

We now consider how \fstar\ is observed to depend on system mass.  We
reproduce the data of LM (including our improved mass estimates
and a correction for the ICL contribution) and GZZ\footnote{Note that the four lowest mass systems shown here were
  excluded from Fig.1 of GZZ, 
because of uncertainty in the gas
mass.  However, the gas mass is irrelevant for our purposes, so we
include these groups here.  Two of the systems, A2405 and APMC020, have
two discrete redshift peaks, and A2405 is a clear superposition of two
systems.}
in Figure~\ref{fig-fstar}.  
The error bars represent one standard deviation from the published
uncertainties on \m500, and we include the correlated uncertainty on
\mstar\ as described in \S~\ref{sec-correrr}.  However, in this case
the direction of the correlated errors is dominated by the fact that
\m500\ appears on both axes.

Both studies
find the stellar fraction decreases with
increasing cluster mass, in qualitative agreement with other work
\citep[e.g.][]{H+00,MH02,Eke-groups3,G+02,RBGMR,cnoc2_ir}. However, it
is clear that the slope of the relationship is different, and the two
datasets are therefore inconsistent 
for the low-mass systems, with $M\lesssim2\times 10^{14}M_\odot$.  
In fact, GZZ 
find a remarkably strong
trend, with a slope $\b=-0.64$, such that the least massive systems in their sample have
\fstar$>0.17$, in excess of even the total {\it baryon} fraction of the
Universe \citep{WMAP3}.  Note that the steep trend is not just driven
by the last four points, however, and the discrepancy exists even if
we ignore these groups.
This is a surprising result and, as we will
show in the following section, potentially poses a challenge to
hierarchical structure growth models.

\section{Cold Dark Matter predictions}\label{sec-models}
\subsection{The progenitor \fstar--\m500\ relation}\label{sec-progmodels}
The data presented in the previous section suggest that \fstar\ decreases 
with increasing mass, above $M\gtrsim 5\times 10^{13}M_\odot$.  Since
CDM theory predicts a robust connection between structures on these
scales, we will attempt to secure a prediction for how steep the 
\fstar--\m500\ relationship can be in this regime.

\begin{figure}
\leavevmode \epsfysize=8cm \epsfbox{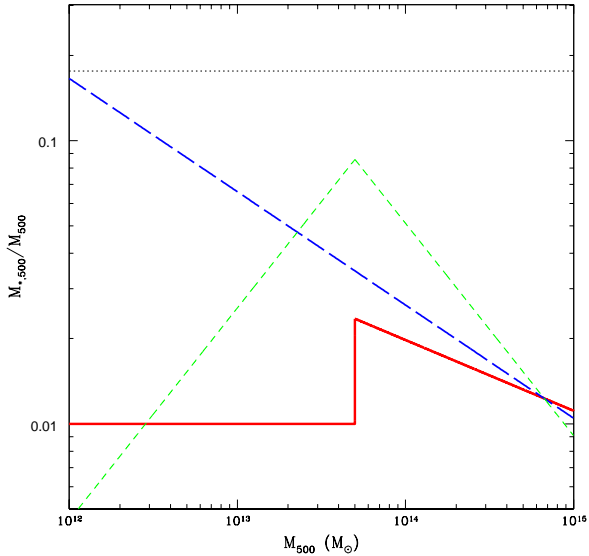}
\caption{Examples of our three assumed models for the dependence of
  \fstar\ on halo mass at $z=0$.  For $M>5\times10^{13}M_\odot$, the {\it Min} model shown
  here (red, solid line) has a slope of $-0.25$, the {\it Extrap} model (blue,
  long-dashed line) has a slope 
of $-0.40$, and the {\it Mirror} model (green, short-dashed line) has a slope
of $-0.75$.  The horizontal, dotted line shows the universal {\it baryon}
fraction from WMAP-3 \citep{WMAP3}.
\label{fig-A}}
\end{figure}

We start by assuming a logarithmic relationship between \fstar\ and \m500,
for systems with $M>5\times10^{13}M_\odot$, at the
present day.  We will then attempt to constrain the slope of this
relation, $b=\b$.  The model is normalized so that 
the most massive clusters, with
\m500$=10^{15}M_\odot$, have \fstar$=0.01$, as motivated by
observations.  
To be conservative we will assume that haloes with $M<10^{11}M_\odot$,
below the resolution limit of our simulations (described in the
following subsection), carry no stars, so \fstar$=0$. For intermediate
masses, $10^{11}<M/M_\odot<5\times10^{13}$, where observational
constraints are most difficult to acquire, we will adopt
three ad-hoc models, which span the full range of realistic behaviour:
\begin{enumerate}
\item{\bf Min:}The stellar fraction is assumed to be constant at a low
  value of \fstar$=0.01$.  This is
the most conservative (non-evolving) model, as it will accommodate the steepest slope
$b$ while maintaining consistency with a low \fstar\ in massive clusters.
\item{\bf Extrap: }The \fstar--\m500\ relation is extrapolated to lower
  masses with the same slope $b$.  However we do not allow it to exceed
  the universal baryon fraction of 17.5 per cent \citep{WMAP3}.
\item{\bf Mirror: }\fstar\ declines with decreasing mass
  below $10^{13}M_\odot$, with slope $+|b|$, mirroring the trend at
  higher masses.  Note that since this is a logarithmic slope, $f_\ast\geq 0.0$ at
  all masses above our resolution limit.
\end{enumerate}
These three models are shown in Figure~\ref{fig-A}, with arbitrary
slopes (for illustration purposes only) $b=-0.25$, $b=-0.40$ and $b=-0.75$ for the {\it Min}, {\it Extrap}
and {\it Mirror} models, respectively.  In all cases $b$ refers to the
slope for  $M>5\times10^{13}M_\odot$, and in practice this slope is a
free parameter that we wish to constrain.  

These models characterise the $z=0$ relationship between \fstar\ and
mass.  This correlation is likely to evolve with
redshift, at a rate that depends on the star formation and mass
accretion history of haloes at a given mass.  
We therefore
consider a generalized model of \fstar:
\begin{equation}\label{eqn-model}
\log{f_\ast}=f_\circ\left(1+z\right)^a+b\left(1+z\right)^c\log{M_{500}},
\end{equation}
where $a$ and $c$ are free parameters that describe the redshift
evolution (in normalization and slope, respectively), and $f_\circ$ is the normalization of the models at $z=0$,
fixed so that \fstar$=0.013$ at $M_{500}=7\times10^{14}M_\odot$.  The present-day
slope $b$ is a free parameter for $M>5\times10^{13}M_\odot$; below
this mass $b$ behaves as described above for the {\it Min}, {\it Extrap}
and {\it Mirror} models.

\subsection{The simulations}

\begin{figure}
\leavevmode \epsfysize=8cm \epsfbox{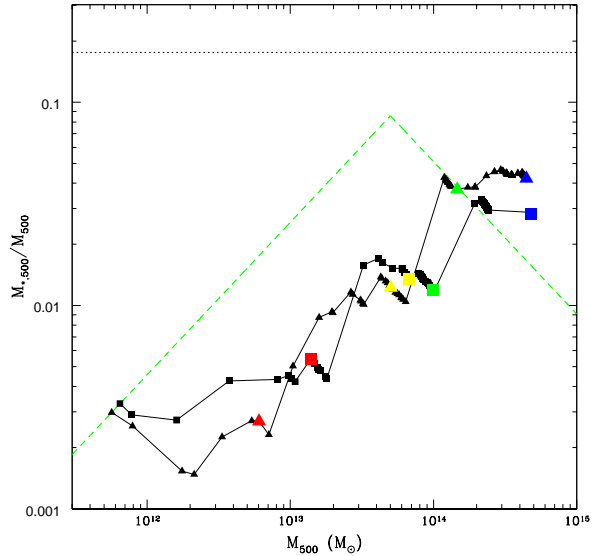}
\caption{Two examples of the growth history of massive clusters from our
  $\Lambda$CDM Pinocchio simulations.  Each point (small triangles
  for one cluster, and small squares for the other) represents a merger
  event above our mass resolution.  The stellar fraction of merging
  haloes are assumed to follow the {\it Mirror} model relation shown
  by the dashed line.  For reference, the large red, yellow, green and
  blue points represent the epochs $z=2.0, 1.0, 0.5$ and $0$.  At early
  times the cluster is growing primarily from sub-resolution accretion;
  because we conservatively assume this accretion carries no stars,
  \fstar\ slowly decreases.  However, most of
  the mass is accreted at late times, through merging with fragments
  that have $M\sim 10^{13}M_\odot$. Therefore, after $z\sim 1$ \fstar\ 
  quickly increases to $\sim 0.03$, and then remains constant at a value
  {\it higher} than the dashed line.  The cluster that accretes more
  of its mass from larger progenitors ends up with a lower \fstar, as expected.
\label{fig-massspectrum}}
\end{figure}
We obtain merger trees generated with the {\it Pinocchio} algorithm \citep{pinocchio}
 to determine the merger history for an ensemble of galaxy
clusters.  The {\it Pinocchio} code has been shown to provide results in
excellent agreement with full N-body simulations (e.g., it reproduces the
mass function of simulated dark matter halos to $\approx 10\%$ accuracy).
We have verified this agreement, for a $\Lambda$CDM universe, by making an
explicit comparison of the predictions of the mean mass growth rate of
massive clusters (and the scatter about the mean) with that of clusters
of similar mass in the publicly available numerical simulation of \citet{Springel05}.
The {\it Pinocchio} code allows us to
efficiently change the mass resolution (to test the sensitivity of the
results to this quantity), to increase the simulation box size (in order
to generate larger samples of the most massive clusters), and to
quickly explore
other cosmological models.  For all the runs presented in this paper, we
adopt a fixed particle mass of $\approx 1.35\times10^{10}M_\odot$, which is
sufficient to follow the growth of the massive ($> 10^{13} M_\odot$)
systems we are interested in.  A ``resolved'' halo corresponds to a system
with a minimum of 10 bound particles.  

We can then compute the final
stellar fraction of a halo of given mass, based on its merger history,
by applying each of the models described in 
\S~\ref{sec-progmodels} to its progenitors. 
As an example, in Figure~\ref{fig-massspectrum} we show the evolution of
\fstar\ for two massive clusters, using a non-evolving ($a=c=0$) {\it Mirror} model
(represented as the dashed line) to assign stellar fractions to each
merging fragment.  Each point represents a merger above our resolution
limit.  At early times, \fstar\ actually declines, because 
most of the mass is being accreted below the resolution
limit, where we assume conservatively that \fstar$=0$.  It is evident
from the distance between adjacent points that most of the mass is
accreted at late times, in low-mass haloes of $M\approx 1$--$10\times 10^{13}M_\odot$; hence \fstar\
rapidly rises to the corresponding value of $\sim 0.03$, and then
remains nearly constant at that level.
The second cluster trajectory we show is deliberately chosen because it
is built from more massive fragments and, as expected, it ends up with
a lower final \fstar.  However, even this cluster has a value of 
\fstar\ that is a factor of two larger than the assumed {\it Mirror}
model.  Therefore this model is internally {\it inconsistent}: as it is
not possible to preferentially remove stars from the final system, we
conclude that \fstar\ is too high in the low mass systems (i.e., $b$ is
too negative) to be consistent with these being the progenitors of
today's clusters.  Since we assume \fstar$=0$ below
our resolution limit, increasing the resolution
(i.e., lowering the mass limit) serves to increase \fstar\ in the final clusters;
apart from this, our results are insensitive to changes in resolution.

\begin{figure}
\leavevmode \epsfxsize=8cm \epsfbox{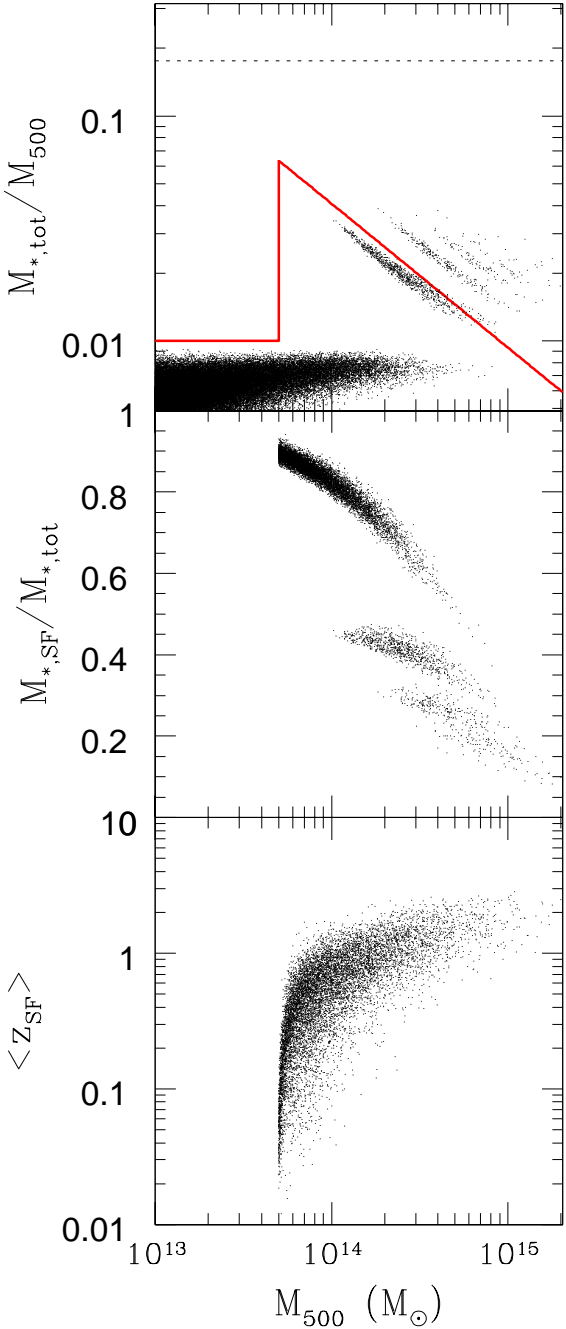}
\caption{ {\bf Top: }  An example of the predicted \fstar\ for an
  ensemble of clusters in a $\Lambda$CDM Universe,
  assuming a non-evolving {\it Min} model with $b=-0.64$ (shown as the
  red, solid line) for the relationship between
  \fstar\ and \m500.  The most
  massive clusters end up with \fstar\ that is much too high,
  indicating an inconsistent model.  In contrast, the lower mass
  clusters lie {\it below} the solid line, because they are built from
  systems with \fstar$\lesssim0.01$.
{\bf Middle: }  This shows the fraction of stellar mass that has to be
  added via in situ star formation to keep haloes on the assumed {\it Min} model (no star
  formation is invoked if \fstar\ is greater than this).  {\bf Bottom:
  } The average redshift at which stars are added to the halo in situ,
  as a function of final mass. In this case, we need to
  add many stars at late times in the lowest mass haloes, to keep
  \fstar\ as high as $0.05$. 
\label{fig-exmodel}}
\end{figure}

\subsection{Example: CDM predictions for a non-evolving, steep
  \fstar--\m500\ relation ($b=-0.64$)}
To demonstrate how our analysis works, we will consider a non-evolving model with
$b=-0.64$, which is the observed $z=0$ slope as measured from the GZZ data.  To be
conservative we will adopt the {\it Min} model prescription for haloes
with $M<5\times10^{13}M_\odot$; this gives the best
chance of obtaining low \fstar\ in the most massive clusters given a
realistic growth history predicted by CDM and a non-evolving
\fstar--\m500 relationship. In the following discussion, the {\it
    model} refers to the assumed \fstar--\m500\ relation that we use to
specify the stellar content of the progenitors for a given cluster.  The {\it prediction} refers
to the \fstar--\m500\ relation that results
when a given model is applied to the progenitors of an ensemble of
clusters with a CDM--specified merger history.  Therefore a
self-consistent model in the context of CDM is one for which the
predicted relation agrees with the model relation.

In the top panel of Figure~\ref{fig-exmodel} we
show the prediction of the final \fstar\ resulting from this particular
model for an ensemble of clusters.   The results are somewhat complex, and we
see the final clusters are grouped into about six ``families''.  Let's
consider first the high mass end, $M\gtrsim5\times10^{14}$, where the
final clusters follow a slope of $b=-0.64$, like the assumed progenitor
model, but with discrete offsets.  The majority of these clusters lie
nearly on the model line.  These are systems which had only one
massive ($M>5\times10^{13}M_\odot$) progenitor, which by construction was
assumed to have a stellar fraction given by the solid line.  A small
contribution to the final mass comes from lower mass systems (with
\fstar$=0.01$) which is why the clusters lie slightly below the solid
line.  The next family of points to the right are those clusters that
formed from exactly two massive progenitors; thus they have similar
values of \fstar\ compared with the first family, but double the final
mass.  Similarly, the clusters found farther to the right are made from
progressively more progenitors with $M>5\times10^{13}M_\odot$.  
Now considering the whole cluster population, the dominant family is
actually comprised of those clusters with \fstar$<0.01$; these are the systems
with zero massive progenitors.   They 
have all been built from low mass haloes, including a significant
contribution from sub-resolution haloes for which \fstar$=0$.

The predicted \fstar--\m500\ correlation is therefore very different from
the assumed model (solid line).  
Since we predict a substantial number of massive clusters with \fstar\ higher
than the solid line, this model is internally inconsistent; 
there is no plausible way to remove stars from the final system, so we
can rule this model out without further
consideration.  

What about the clusters that are predicted to have a {\it lower}
\fstar\ than the assumed model?  Unfortunately, these do not represent
such a clear failure of the model, since one could always assume that
systems form enough stars in situ, {\it after} a merger event, to move them up to
the appropriate value of \fstar\ for their mass\footnote{Note therefore the
subtle distinction, that a non-evolving \fstar--\m500\
model is not synonymous with a lack of star formation.}.  The middle panel of Figure~\ref{fig-exmodel} 
shows (for clusters with $M>5\times10^{13}M_\odot$) the additional
fraction of stars that must be formed in this way to ensure a
consistent model.  The discontinuity in the
model means that clusters just above this threshold must form most of
their stars in situ, since their progenitors all have \fstar$\leq0.01$.
Moreover, most of this in situ star formation has to occur at late
times, since for most of their existence these systems will have had
$M<5\times10^{13}M_\odot$, and they have only recently crossed this threshold.
We deal with this in the following way.  After each merger event, we
add sufficient stars, via in situ star formation, to bring the cluster
back up to the model (solid line).  No stars are added to (or removed
from) systems with larger \fstar\ than the model.  We can then track
the average redshift at which those additional stars were 
added, and this is shown in the bottom panel of
Figure~\ref{fig-exmodel}. At $M\sim5\times10^{13}M_\odot$, most of the
stars must be formed
very recently, as argued above. This large amount of recent star
formation in groups is not supported by observations, a point on which
we will elaborate in \S~\ref{sec-constraints}; so this is another
indication that the assumed model is unphysical.  For the most massive clusters, we only
require that $\sim 10$ per cent of the stars are formed in situ, and
this at $z>1$, which is much more reasonable. 

This fairly complex behaviour nonetheless reflects a consistent trend in most of our
models.  The hierarchical merging process always tends to produce final
systems with \fstar\ that is {\it nearly independent of halo mass}.
Thus, for models with a steep $b$, the most massive clusters end up
with \fstar\ that is too high (and thus inconsistent), while the least
massive end up with \fstar\ that is much too low, and therefore
require a lot of recent, in situ star formation.  This occurs because the mass
accretion histories of clusters over this limited mass range are not
very different -- they are built from similar-mass haloes over a similar
time -- resulting in a similar final \fstar.  Thus we can
immediately see from this simple example that CDM will
prefer a value of $b$ that is much closer to $0$ than the GZZ
observations suggest (for non-evolving models).

We have neglected any discussion of the gaseous component of clusters;
not only is it dynamically of minor importance, making up $< 20$ per
cent of the cluster mass \citep[e.g.][]{Allen+04}, but it is expected that major mergers are
most likely to {\it remove} gas from the dark matter
\citep[e.g.][]{bullet}, and hence further increase \fstar.  However, it
is also possible that low-mass clusters and groups are relatively
deficient in gas \citep[e.g.][]{ArEv99,Vik+06}, perhaps because it has been
preheated \citep[e.g.][]{Babul2,McCarthy-MT}; if this gas is accreted later during
the hierarchical growth of structure it could cause \fstar\ to
decrease.  Specifically, if we take the extreme assumption that systems with
$M=10^{13}M_\odot$ have {\it no} associated gas, then this effect alone
would lead only to $b\approx -0.04$, assuming that systems with
$M=10^{15}M_\odot$ have accreted their full complement of gas. 

This model also implicitly assumes that the radial distribution of
stars traces that of the dark matter and, in particular, is not altered
by the merger process.  This is a reasonable assumption, as there is no
evidence that the stellar fraction is a strong function of radius
outside the very centre of clusters, and the stellar light distribution
is usually found to be well modelled by a \citet{NFW} profile with a
reasonable concentration parameter \citep[e.g.][]{LMS} and comparable to the total
mass distribution \citep{CYE,Muzzin07}, even for relatively low mass
systems \citep[e.g.][]{Sheldon+07}.  Furthermore, while the merging process can
greatly distort the relative distribution of gas and dark matter, it is
much less likely to separate the stars from the dark 
matter \citep[e.g.][]{bullet}.  

\subsection{Constraints}\label{sec-constraints}
The previous example demonstrates that there are two aspects of our predictions that we can use to choose
acceptable models.  If many systems end up with \fstar\ greater than
assumed in the model, we can confidently rule it out; there is no reasonable way
to reduce the number of stars, so the model is internally
inconsistent. 
However, a model could also be deemed
unreasonable if it requires that most of the stars are formed too recently.
This second constraint is less robust because we require
guidance from observations.  However, we will show
below that we can afford to be quite conservative.

\begin{figure}
\leavevmode \epsfysize=8cm \epsfbox{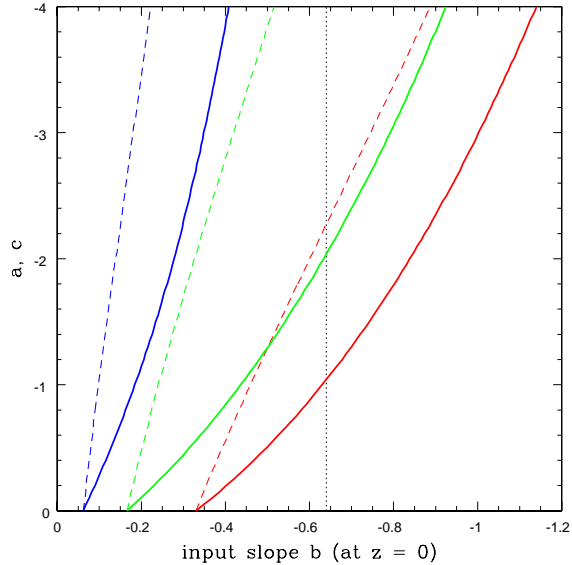}
\caption{The maximum slope $b=\b$ at the present day that results in a
  consistent model with \fstar$\leq0.01$ in the most massive clusters.
  The line colours correspond to the different models shown in
  Figure~\ref{fig-A}, which make different assumptions about the
  behaviour of \fstar\ for haloes with $M<5\times10^{13}M_\odot$.
  Dashed lines correspond to evolution in the slope (parameter $a$ in Equation~\ref{eqn-model})
  while solid lines represent evolution in the normalization (parameter
  $c$). Parameter space to the right of any curve
  is excluded, as it would lead to clusters with too many stars today.
  The vertical, dotted line represents the slope measured from the GZZ data.
\label{fig-bmax}}
\end{figure}

In Figure~\ref{fig-bmax} we show, for a range of evolution parameters
$a$ and $c$, the most negative value of $b$ that yields internally consistent clusters, i.e. those for which
the predicted \fstar\ on average lies on or below the assumed \fstar--\m500\
model.   In principle the evolution could be positive, such that \fstar\ at
a given mass scale is greater at high redshift than it is today;
for example, if efficient star formation on some scale at $z=1$ is
followed by a long period of quiescent accretion of lower mass haloes,
in which star formation has been inefficient.  However, in practice,
only negative evolution puts interesting constraints on the slope $b$,
so we focus our attention on that regime.

For each model, represented by a curved line, the region
to the right of the plot is excluded.  For non-evolving models,
$a=c=0$, all models show $b>-0.33$.  This means that, given
\fstar$=0.01$ in the most massive clusters, groups with
$M\approx5\times10^{13}M_\odot$ must have \fstar$<0.025$.  This is
considerably shallower than the slope found by GZZ, shown as the
vertical, dotted line.

If we allow the $z=0$ relation to evolve strongly, we can weaken our
constraints.   This is because the groups that
merge to form clusters will have done so at a higher redshift than the
groups we are looking at today (recall Figure~\ref{fig-massspectrum}). 
Figure~\ref{fig-bmax} shows that an arbitrarily steep
\fstar--\m500\ relation at the present day can be accommodated if it is
allowed to evolve strongly enough.  However, this success comes at a high
price. In hierarchical models, groups with lower mass form at even
{\it higher} redshifts than the clusters, so they will have been
built from systems with even lower values of \fstar.  The natural
prediction of these 
strongly evolving models is therefore that \fstar\ at $z=0$ will be low not
just in clusters, but in {\it all} haloes.  Thus, in order for our model
to be consistent, we require that a great deal of stars form in situ in these
lower mass haloes, {\it after} most of the clusters have been assembled,
which, for a
$\Lambda$CDM Universe, is $z\approx 0.3$ (see Figure~\ref{fig-cosmo}).
The more steeply we assume \fstar\ evolves,
the more recently those stars must have been created.  

\begin{figure}
\leavevmode \epsfysize=8cm \epsfbox{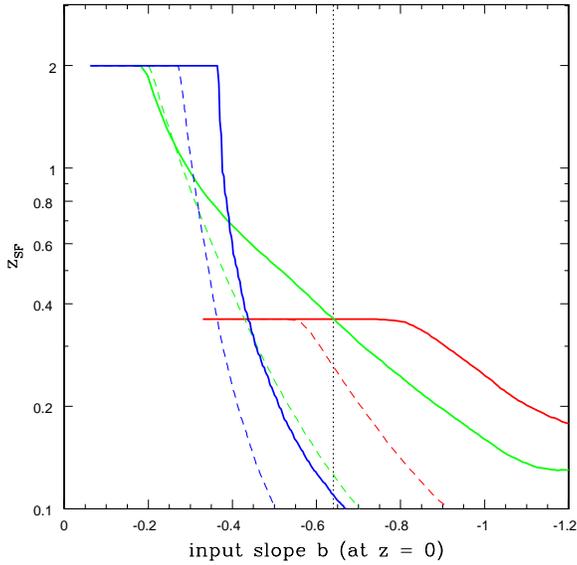}
\caption{The median redshift at which stars must have {\it formed} in
  haloes with $5\times10^{13}<M/M_\odot<10^{14}$, to remain consistent
  with each model, as a function of the maximum allowable slope $b$ at
  the present day, from Figure~\ref{fig-bmax}.  The line styles
  correspond to those in Figure~\ref{fig-bmax}.  For each model, more
  negative values of $b$ imply greater evolution in the \fstar--\m500\ relation.  
  The vertical, dotted line represents the slope measured from the GZZ data.
\label{fig-zmed}}
\end{figure}

We quantify this in
Figure~\ref{fig-zmed}.  For each halo at $z=0$, we assume that any
stars that were generated via in situ
star formation formed at the
first redshift that any merger yields a stellar fraction below the
assumed value\footnote{In reality, the stars could form anytime between then and the
next merger.  This means our results are conservative.}.  Stars that were
accreted through mergers are conservatively assumed to have formed at $z=2$.
We use this to compute, for each halo, the redshift at which half the
stars were formed in situ, $z_{\rm SF}$.  Figure~\ref{fig-zmed} shows the
median $z_{\rm SF}$ for haloes with $5\times10^{13}<M/M_\odot<10^{14}$
(i.e. at the low mass end of our models), as a function of the maximum $z=0$
slope $b$ allowed by each model, from Figure~\ref{fig-bmax}.  Recall
that more negative values of $b$ imply steeper evolution.  As described
above, the success of strongly-evolving models to match the low \fstar\
in today's clusters comes at the expense of requiring substantial
recent star formation in groups.  

Although it is clear that galaxy cluster populations are old and
passively evolving, with
$z_f\gg2$ \citep[e.g.][]{BLE,Roberto99,Roberto07,Finn05,Nelan,LMGS,Muzzin07},
there are fewer constraints on lower mass groups.   However, globally we
know that the total mass in stars has at most doubled since
$z=1$ \citep[e.g.][]{Dickinson03,Bell+03,Drory+04,GH05,Pozzetti+07}. 
Relative to the global population, at $z\lesssim 0.5$ groups are
known to be dominated by galaxies with early-morphological types and
little or no star formation \citep[e.g.][]{dlR01,Tran,lowlx-spectra,2dfsdss,cnoc2_ir,Weinmann+06,CNOC2_groupsI,JMLF,McGee}; therefore we would
expect them to have formed half their stars well before $z=1$.  
Recently, \citet{Brough+07}
have made a detailed analysis of the brightest galaxies (which,
together with the ICL, GZZ claim dominate the stellar mass) in three X-ray
groups.  Two of these have luminosity--weighted ages $>10.5$ Gyr ($1\sigma$) limit,
corresponding to a formation redshift of $z_f=2$; the youngest has an age
limit of $>6.9$ Gyr, or $z_f=0.75$.  

We therefore consider that
$z_{\rm SF}\gtrsim 1$ (a lookback time of 8 Gyr) is a very reasonable lower limit for the redshift at
which groups ($5\times10^{13}<M/M_\odot<10^{14}$) have formed most of their stars.  From
Figure~\ref{fig-zmed} this implies a lower limit of $b>-0.35$, only
slightly steeper than the non-evolving model constraints.
To accommodate a slope $b=-0.64$, as observed by GZZ, would require
that at least half the stars in groups with
$5\times10^{13}<M/M_\odot<10^{14}$ formed since $z=0.35$, which is 
very unlikely given the above observations.  

\subsection{Ab-initio models}\label{sec-bower}
\begin{figure}
\leavevmode \epsfysize=8cm \epsfbox{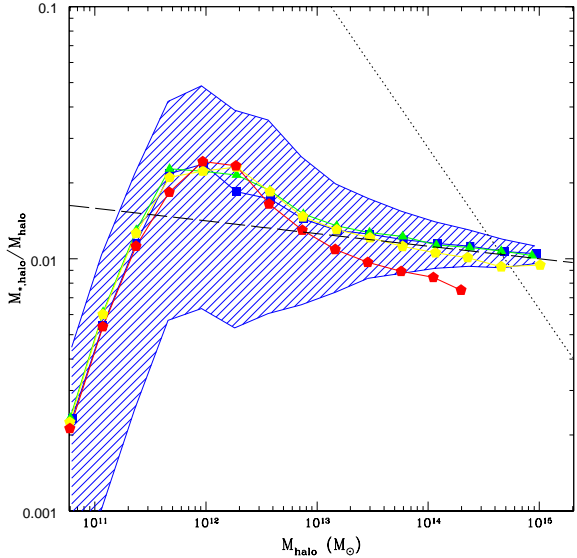}
\caption{The predicted stellar fraction as a function of halo mass
  (approximately $M_{200}$),
  from the galaxy formation models of \citet{bower06}.  The points
  represent binned averages at different redshifts: blue, green,
  yellow, and red points correspond to $z= 0$, 0.5, 1.0, and 2.0,
  respectively.  The blue shaded region corresponds to the 10th and
90th percentiles of the $z=0$ distribution. The dashed line is an
approximate representation of the $z\leq 1.0$ model data for $M_{500}>
10^{13}M_\odot$, and has a slope of $b=-0.05$.  The  dotted line represents the
approximate slope ($b=-0.64$) of the GZZ 
data, with arbitrary normalization.
\label{fig-bower}}
\end{figure}

Instead of assuming a correlation between \fstar\ and \m500, an
alternative approach is to investigate more complex, ab-initio models that
include prescriptions for star formation and feedback.  There are
many such models currently available, with generally similar recipes
\citep[e.g.][]{DL+05,Croton05,bower06}.  Here we consider the
publicly available\footnote{http://www.icc.dur.ac.uk/} predictions of the \citet{bower06} model.  Figure ~\ref{fig-bower}
shows the predicted \fstar\ distribution as a function of halo mass, 
for all systems in the parent simulation, at redshifts between $z=0$ and $z=2$.
For halo mass, we use a friends-of-friends mass with linking length
$b=0.2$, which corresponds approximately to $M_{200}$, with a scatter of
about 15 per cent.  For a typical cluster halo, \m500$\approx 0.65 M_{200}$.
There is almost no mass dependence of \fstar\ for 
$M>10^{13}M_\odot$, and very
little evolution in the relation.  Above $M=10^{13}M_\odot$, the model
at all redshifts satisfies approximately $\b=-0.05$.
This prediction
for a nearly constant \fstar\ is also in good agreement with
cosmological hydrodynamic simulations that include cooling and
feedback physics \citep[e.g.][]{Borgani04,Kay+07}, although those
simulations tend to  predict a higher overall value of \fstar.  

This prediction for a nearly constant \fstar\ on these scales is not
surprising, given that massive groups and clusters are both built from
haloes with a similar mass distribution, and over a similar timescale.
Although our arguments from the previous section demonstrate that
CDM {\it could} support a \fstar-\m500\ relation as steep as $b\approx
-0.35$ at
the present day, this is only true under the most conservative, and
probably unrealistic, assumptions about the value of \fstar\ in haloes with 
\m500$<5\times10^{13}M_\odot$ and how this evolves with redshift.

\subsection{Other cosmologies}\label{sec-cosmo}
\begin{figure}
\leavevmode \epsfysize=8cm \epsfbox{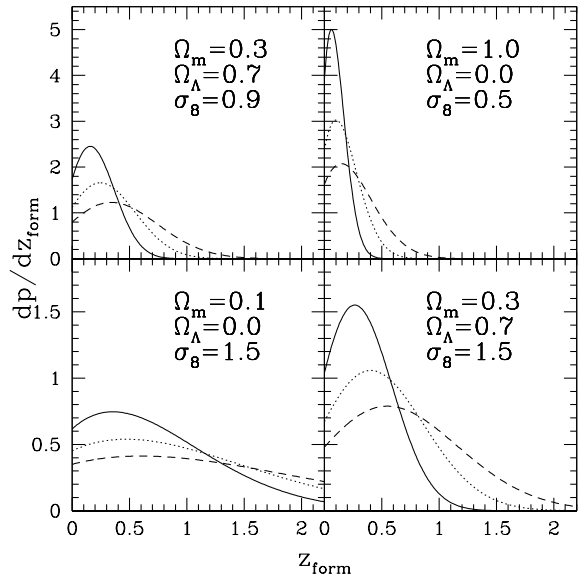}
\caption{The probability distribution of formation redshifts, defined
  as the redshift where 75 per cent of the mass is assembled, for
  clusters with virial masses $M=10^{15}M_\odot$ (solid line), $M=10^{14}M_\odot$
  (dotted line), and $M=10^{13}M_\odot$ (dashed line).  The
  calculations are based
  on the analytic model of \citet{LC93}.  Different panels show the
  results for different cosmological parameters, as indicated.  The
  highest formation redshifts are achieved in low-$\Omega_m$ or
  high-$\sigma_8$ models.  But in all cases, low-mass haloes form at
  substantially higher redshift than high-mass haloes.  Thus,
  cosmological parameters have little effect on the generic prediction
  of hierarchical models, that $b=\b$ as measured at the present day
  must be flat, $b>-0.35$.
\label{fig-cosmo}}
\end{figure}

We can hope to put constraints on the cold dark matter model in
general, regardless of the specific cosmology, since they are all
hierarchical in nature.  Different cosmological parameters will result
in different rates of formation for haloes of a given mass. We showed
in \S~\ref{sec-constraints} that one can maintain a 
low \fstar\ in the most massive clusters today, whatever the local
\fstar--\m500\ relation, as long as this relation evolves strongly with
redshift.  Clearly one could achieve this same result with weaker
evolution, in a cosmology where structure forms earlier.  This is true
of an open Universe, with $\Omega_m<1$ and $\Omega_\Lambda=0$
\citep[e.g.][]{vdB02}; alternatively the epoch of structure formation
can be pushed to higher redshift if the normalization of the power
spectrum, $\sigma_8$, is increased \citep[e.g.][]{LC93}.  This is
illustrated in Figure~\ref{fig-cosmo}, where we use the \citet{LC93}
model to calculate the probability distribution of the formation
redshift, defined as the redshift where 75 per cent of the final mass
is in place \citep[see details in][]{entropy}.  Distributions are shown
for haloes with final masses of $10^{13}M_\odot$,  $10^{14}M_\odot$,
and $10^{15}M_\odot$, for a range of cosmological parameters.  As
expected, the highest formation redshifts are obtained in a
low-$\Omega_m$ or high-$\sigma_8$ Universe.

However, in all cases, lower-mass haloes on average form at even {\it
  higher} redshifts; this of course is the well-known behaviour of cold
dark matter models.  Therefore, in a low-$\Omega_m$ or high-$\sigma_8$
Universe, galaxy groups had most of their mass in place at even higher
redshift than the clusters, and \fstar\ today should also be
lower, requiring considerable late-epoch star formation in groups to
retain consistency with a steep relation $b\ll0$.  This trade-off means
that our constraint on $b$ is unlikely to be changed in different cosmologies.

We have rerun our {\it Pinocchio} simulations, for an
Einstein De-Sitter Universe ($\Omega_m=1$, $\Omega_\Lambda=0$, $\sigma_8=0.5$) and an
open Universe ($\Omega_m=0.1$, $\Omega_\Lambda=0$, $\sigma_8=0.9$).  To be conservative
we adopt the {\it Min} model for the \fstar--\m500\ relation at $z=0$,
and allow the normalization or slope to evolve rapidly, as $(1+z)^2$.
The minimum value of $b$ in this model, that is consistent with
\fstar$=0.01$ in the most massive systems and ensures that at least
half of the stars formed at $z>1$, is still $b\approx -0.3$ for the EdS model,
but could be relaxed to $b\approx -0.4$ for the OCDM cosmology, only slightly
steeper than the constraint for our
default $\Lambda$CDM case.  Thus, our conclusions
are nearly independent of cosmological parameters, as expected.

\section{Discussion}\label{sec-test}
\subsection{Observational Uncertainties}\label{sec-obserrs}
As published, the GZZ 
data show stellar fractions reaching as
high as 30\% in the lowest mass systems (which, we note, have poorly
calibrated \m500), well above the WMAP
constraints on the global baryon fraction \citep{WMAP,WMAP3}.  If this
is confirmed to be representative of systems in this mass range, it will rule out hierarchical structure formation
models.  The shallower relation between \fstar\ and \m500\ as found by LM, however,
does appear to be consistent with such models. In this section we
will look carefully at some of the biases and
uncertainties involved in the analysis.

\begin{figure}
\leavevmode \epsfysize=8cm \epsfbox{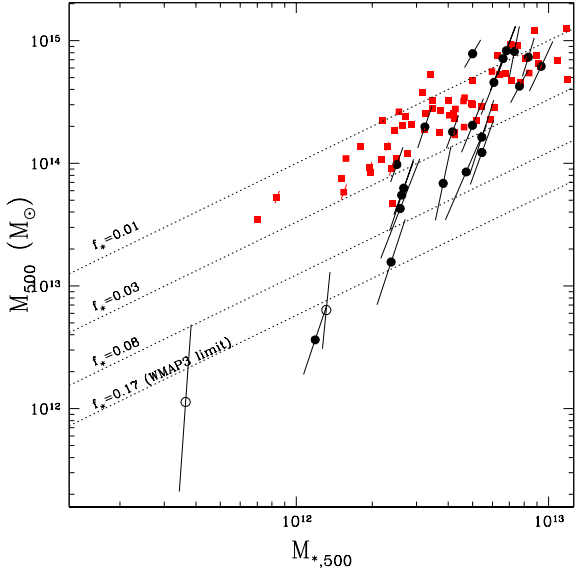}
\caption{The total mass within \r500\ is shown as a function of the
  total stellar mass (including intracluster light) within the same
  radius, for the LM sample (squares) and the GZZ
  sample (circles).  The two open circles
  represent clusters  A2405 and APMC020, which are systems strongly
  affected by line-of-sight structure.  The dashed lines show lines of constant stellar
  fraction, as indicated.  Error bars reflect the uncertainty on
  \m500, and the tilt shows how this is correlated with \mstar.
\label{fig-lstar}}
\end{figure}

One of the most important differences between these two
surveys is in the measurement of total mass. GZZ 
derive masses from the velocity
dispersions, which have significant uncertainties themselves, and
translate into relative errors three times larger when converted to
\m500. Moreover, the strong correlation between errors on \m500\ and
\mstar, as discussed in \S~\ref{sec-correrr}, further complicates matters.  To
illustrate this better we replot the data in Figure~\ref{fig-lstar},
this time showing \m500\ as a function of \mstar, so that the tilt in
the error bar now reflects the degree of correlation between these two
measurements. 

It is evident that the error bars have the same ``slope'' as the GZZ
data itself, suggesting that the steepness of their relation is at
least partly due to this correlation.
However, this may not be the whole story.  Accounting for the
correlation on the errors, we can compute how likely it would be to
find as many systems with \fstar$>0.05$ as GZZ do (seven), if they all actually had
\fstar$=0.02$ (consistent with LM), and random uncertainties on
$\sigma$ scatter the observations.  This probability is less than 0.1
per cent. The uncertainties on \m500\ do not include the scatter in the
$\sigma$--\m500\ relation from which it is derived; %this is at least 5
however, even increasing the error bars
on $\sigma$ by 50 per cent, we find that we would only expect to find
seven systems with \fstar$>0.05$ six per cent of the time.  It is clear
from Figure~\ref{fig-lstar} why this is; the GZZ data lie
systematically below those of LM; while any one point may be discrepant
by only one or two standard deviations, the difference between the two
relations is much more significant.  If the source of the discrepancy
lies in the total masses, then, it implies that these masses are
systematically underestimated for \m500$\lesssim10^{14}$.  One possibility
is that the velocity dispersions themselves are underestimated;
however, this seems unlikely, as several authors have
found that $\sigma$ is quite robust for systems with at least
20 members \citep[e.g.][]{ZM98,Borgani+99,HEHY}. The other, more likely
explanation is that the adopted calibration between $\sigma$ and \m500\
is incorrect at low masses.  In particular, the four groups with the
lowest $\sigma$ require an extrapolation of this relationship, which
may well be unreliable.  Ignoring these four
clusters, the remaining GZZ data are statistically consistent with a stellar fraction
of $0.02$ that is {\it independent} of cluster mass, and an apparently steep
slope that is due entirely to the correlated error bars.  

\begin{figure}
\leavevmode \epsfysize=8cm \epsfbox{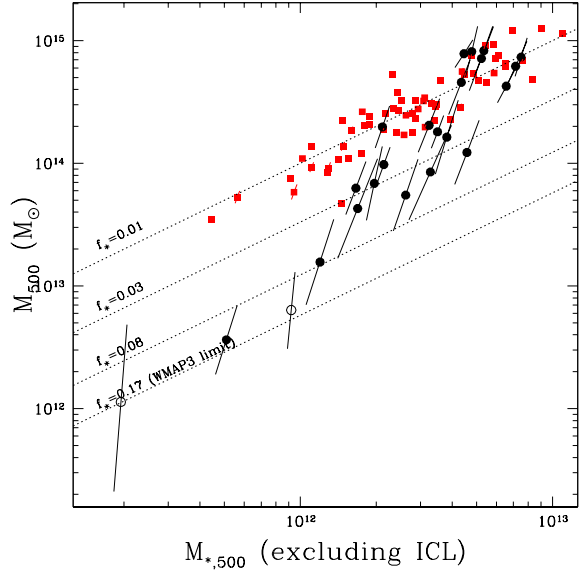}
\caption{As Figure~\ref{fig-lstar}, but where we have excluded the ICL
  contribution to the stellar light, for both surveys.  
\label{fig-lstar_noicl}}
\end{figure}

There are other possible sources of systematic error in the data, but
none of them seem likely to account for the discrepancy.  The most
obvious place to look is the intracluster light, since this was not
directly measured by LM.  In fact, the claim by
GZZ that the ICL fraction is such a strong function of mass is not
readily apparent in the recent measurements
from \citet{Zibetti}, based on ensemble averages of SDSS clusters.
Moreover, numerical simulations generally find that the ICL component
should actually be {\it less} important in groups, relative to
clusters \citep[e.g.][]{Murante+04,Murante+07}.
However, even though GZZ 
claim the ICL doubles the stellar
mass in the smallest observed systems, this is still not enough to
account for the discrepancy with LM.  In Figure~\ref{fig-lstar_noicl} we replot the data shown
in Figure~\ref{fig-lstar}, but excluding the ICL component from both
samples.  There is still a significant discrepancy between them, and
the lowest mass systems in GZZ 
have stellar fractions $>10$\%.  In particular, if the ICL is ignored,
then LM observe a nearly constant \fstar\ over all masses, not only
consistent with our theoretical bounds but also in good quantitative
agreement with ab-initio models \citep[e.g.][]{bower06}.  On the other
hand, GZZ still predict a strong mass dependence of the fractional galaxy light alone.

The dominant systematic uncertainty in the stellar mass measurements
are likely in the stellar mass-to-light ratios. The absolute, average, $M/L_I$
adopted is of little consequence, but the relative difference between
$M/L_I$ and $M/L_K$, and any trend in these values with total mass, is
relevant.  The relative value of  $M/L_I$ and $M/L_K$, and hence the
relative normalization of the data in Figs~\ref{fig-fstar}, \ref{fig-lstar} and
\ref{fig-lstar_noicl}, depends on our assumption that the average
galaxy colour is $I-K=2.0$ in these samples.  A bluer population would
act to decrease the GZZ stellar masses, relative to LM, and thus
decrease the systematic offset observed in these figures.  However, no
reasonable colour will reconcile the four lowest-mass GZZ groups with
those of LM, and the remainder are already consistent with LM (and a
constant stellar fraction) given the error bars.  Of more interest is
the possibility that $M/L$ varies with system mass.  If lower-mass
clusters have systematically younger stellar
populations they will have lower average $M/L$, and if unaccounted for
this would lead to an artificially steep slope $b$ (and would
also compromise some of the conclusions in GZZ).  However, to explain
the discrepancy with the LM data would require that the low-mass
clusters of GZZ are systematically bluer than those of LM, while the
massive clusters are similar.  
It seems unlikely that either of these biasses are important, since 
GZZ find that
most of the stars in these low-mass systems are in the ICL and BCG
component, and the BCG at least is not likely to be much younger than in
more massive clusters \citep[e.g.][]{Brough+07}.  
Moreover, the average stellar M/L in the $I$
band is unlikely to vary by more than a factor of about two, which
is insufficient to reconcile the steep slope $b$ observed by GZZ with the model
predictions.  
Nonetheless, deep infrared images of these 
systems would be very useful.

Another concern might be the
contribution to the stellar light from faint, unresolved galaxies;  however, both
GZZ 
and LM assume fairly steep, mass-independent faint end slopes when
extrapolating the galaxy luminosity function ($\alpha=-1.21$ and
$\alpha=-1.1$, respectively), so this cannot contribute to the
difference.  Finally, there is a possible selection effect as GZZ select groups to have a
dominant galaxy, and this may bias them toward systems with
particularly high stellar fractions.  Observations of a more
representative sample would be valuable, especially if there are
sufficient redshifts to robustly identify cluster members.  However, in
the \citet{bower06} models we considered in \S~\ref{sec-bower}, {\it
  none} of the systems in the relevant mass range are found to have stellar fractions as high as
even 5\%.  If these models provide an accurate picture of the local
Universe in this respect, then no possible selection bias would lead to
the high values of \fstar\ observed by GZZ.

\subsection{The consequences}
If the strong trend observed by GZZ 
is confirmed, and found to
be typical of a mass-limited sample of groups, what are the theoretical
consequences?  We have shown that such groups cannot be the progenitors
of today's clusters, even if they have built up their high stellar
fraction quite recently.  One possible implication would be that today's
clusters have grown very little in mass since at least $z=1$, while
groups have assembled more recently.  This inherently non-hierarchical
model would require a lot of
suppression of power on group scales; qualitatively this is the
behaviour of warm dark matter models, although these generally suppress
structure on much smaller scales \citep[e.g.][]{A-R+01,BOT01}.  Another possibility is
that a large fraction of massive dark matter haloes have {\it no} stars
associated with them, or at least not enough to be detectable.  There
is some preliminary evidence for such dark clusters \citep{vdL+06,MHBBC}.  However,
for
this to be the solution, the most massive clusters would have to have
accreted $\sim 80$\% of their mass from such objects, while groups
still accrete all their mass in haloes with \fstar$>0$.  This seems
quite unlikely, but should be testable in forthcoming weak-lensing
surveys.  

On the other hand, if the dynamical masses of the groups in the GZZ sample are
underestimated, this raises another theoretical problem.  A
particularly nice result from GZZ is that (ignoring the four lowest
mass clusters) their data give a full account of the baryons: the sum
of the stellar mass and the expected gas mass (unfortunately, not
directly measured) is comparable to the total baryon fraction in the
Universe, and independent of system mass.  This is an attractive 
explanation for the observation that galaxy groups are deficient in
X-ray gas \citep[e.g.][]{ArEv99,Vik+06}: the missing gas has cooled to
form stars.   If GZZ have overestimated \fstar, however, this
explanation will no longer be viable, and the low
gas fractions of groups would likely imply that some powerful form of heating
has expelled a large fraction of the gas beyond \r500.  However, this
heating must not be too strong, so that the most massive systems are
still able to retain all their gas.  This requires something of a
delicate balance.

\section{Conclusions}
The stellar fraction \fstar\ as a function of cluster mass \m500\ is an important test
of hierarchical structure formation models.  We find that such models
can make a robust, falsifiable prediction that the power-law slope
relating these two quantities is $\b>-0.35$, for systems with
$M_{500}>5\times10^{13}M_\odot$. A steeper slope can only be accommodated
if the \fstar--$M_{500}$ relation evolves strongly, such that
galaxy groups formed most of their stars in situ since $z=1.0$, which is not
supported by observations.
Since the most massive
clusters today (with $M_{500}\sim10^{15}M_\odot$) are robustly measured
to have \fstar$\sim0.01$, hierarchical models
therefore require that galaxy groups (with $M_{500}\sim5\times 10^{13}M_\odot$)
have \fstar$<0.03$.  This constraint is a conservative limit;
ab-initio models predict a much flatter relationship with $b>-0.1$ \citep{bower06}.

Recent observations by \citet{GZZ} 
appear to conflict with this prediction.  In particular,
their data follow a power-law relation over almost two orders of
magnitude in mass, with $b=-0.64$, and  the
lowest-mass systems in their sample have stellar fractions of 30 per cent,
exceeding even the limits on the baryon fraction from WMAP.  If
confirmed, these observations definitively rule out hierarchical structure
formation.  $K-$band observations from \citet{LM04}, on the other hand,
are just consistent with the model constraints, but may still be inconsistent
with the ab-initio model of \citet{bower06} if the ICL contribution is
as dominant in groups as claimed by \citet{GZZ}.  These data do not extend to
such low mass systems, nor are their data deep enough to directly
measure the important ICL contribution.  More observations are needed to resolve
the discrepancy between these two studies, and to thereby test the viability of
cold dark matter models.  Particularly valuable would be X-ray images
(and temperature maps)
of the \citet{GZZ} sample, to measure the gas content and total mass, and deep $I$ or near-infrared images of the
\citet{LM04} systems to directly measure the ICL component.
 
We end by noting however that the four lowest-mass systems in the
Gonzalez et al.
sample have poorly calibrated masses.  If we accept that their masses
are underestimated by a factor of ten, then the
remainder of their data are consistent with a universal stellar fraction
of a few percent, and the apparent correlation with system mass is due
to the correlated error bars on dynamical and stellar
masses. Furthermore, removing these four systems greatly reduces the
significance of their claim that the intracluster light fraction is
a strong function of mass and, in this case, the \citet{LM04} data (which have
much smaller error bars) are also consistent with a constant stellar
fraction \fstar$\approx 0.01$. 
This
conclusion challenges the claim by GZZ that the stellar component of
galaxy groups is dominated by the BCG and ICL, and that the low gas
fractions in these groups is attributable to an increased stellar fraction.

\section{Acknowledgments}\label{sec-akn}We thank the referee, Anthony
Gonzalez, for his thorough report, which led to
substantial improvements in this paper.
This research was supported by the Natural Sciences and Engineering
Research Council of Canada, through a Discovery Grant to M. Balogh.  IGM acknowledges support from a NSERC Postdoctoral
Fellowship, and thanks Tom Theuns for his assistance with
the use of the {\it Pinocchio} software.  VRE acknowledges support from the Royal Society for a University Research
Fellowship.
\bibliography{ms}
\end{document}